\documentclass{dcds2}
\usepackage[dvips]{graphicx}

\def\ppages{1--10}

\theoremstyle{plain}

\theoremstyle{definition}

\newcommand{\Hamil}{{\mathcal H}}

\newcommand{\Elinnd}{E_h}

\newcommand{\Elinkd}{E_{h,\kdisc}}

\newcommand{\kdisc}{k}
\newcommand{\kdisca}{\kdisc_1}
\newcommand{\kdiscb}{\kdisc_2}
\newcommand{\kdiscc}{\kdisc_3}
\newcommand{\Ifour}{V}
\newcommand{\sgn}{\,\mathrm{sgn}\,}
\newcommand{\kknee}{k_{\mathrm{knee}}}
\newcommand{\neff}{n_{\mathrm{eff}}}
\title[Stages of Energy Transfer in the FPU Model]
 {Stages of Energy Transfer in the FPU Model}

 \email{biellj@rpi.edu}

\author[Joseph A. Biello, Peter R. Kramer, and Yuri Lvov]{}

\def\cal#1{{\fam2#1}}

\begin{document}
\maketitle

\centerline{\scshape Joseph A. Biello, Peter R. Kramer, and Yuri V. Lvov }
\medskip

{\footnotesize \centerline{} 
\centerline{Department of Mathematical Sciences}
\centerline{Rensselaer Polytechnic Institute}
 \centerline{301 Amos Eaton Hall}
 \centerline{110 8th Street}
 \centerline{Troy, NY 12180, USA}
 }

\begin{quote}{\normalfont\fontsize{8}{10}\selectfont
{\bfseries Abstract.}
The (alpha) version of the Fermi-Pasta-Ulam is revisited through
direct numerical simulations and an application of weak turbulence
theory.  The energy spectrum, initialized with a large scale
excitation, is traced through a series of distinct qualitative phases
\emph{en route} to eventual equipartition.  Weak turbulence theory is
applied in an attempt to provide an effective quantitative description
of the evolution of the energy spectrum.  Some scaling
predictions are well-confirmed by the numerical simulations.
\par}
\end{quote}

\section{Introduction}
One of the very first uses of electronic computing machines was Fermi,
Pasta, and Ulam's simulation of wave propagation in a weakly nonlinear
lattice model~\cite{ef:snp}.  They were expecting to observe that
the weak
coupling of the normal modes of the system would induce a
redistribution of energy from an initial large-scale excitation to an
equal sharing (equipartition) of energy among all normal modes after
some time.  As is well known, they were instead surprised to see the
system display regular behavior characteristic of integrable systems,
with the initial state recurring on a rather short time scale.  This
discovery shifted attention to its explanation and
ramifications~\cite{ei:nwsc,acn:smp} for several decades.
  In the last two decades,
however, the
Fermi-Pasta-Ulam (FPU) model has once again been utilized as a test model for
numerically 
illustrating and exploring standard concepts in statistical
mechanics~\cite{lc:fpupr}.  The peculiar near-integrable 
behavior observed by Fermi, Pasta, and Ulam is
characteristic of their model only for systems which are sufficiently
small in size and energy.  There exist well-defined regimes for which
the FPU model is weakly nonlinear but stochastic 
\cite{mcs:psgst,rl:etnlh,pp:esfpu}, 
and it is in these regimes that one can
hope to connect the outcome of direct numerical
simulations with statistical mechanical concepts~\cite{lc:fpupr}, such
as relaxation to equipartition~\cite{lc:fpupr,jdl:ftetl,vvm:cbfce,pp:swtsr}, entropy production~\cite{lc:fpupr,jdl:ueeoc,jdl:ftetl,cyl:wodls,vvm:cbfce,pp:swtsr}, chaos as manifested by positive
Lyapunov exponents~\cite{ca:qhpsf,lc:fpupr,td:memle,pp:esfpu,dls:lecfp}, universal behavior of statistical functions~\cite{jdl:ueeoc,jdl:ftetl,cyl:wodls,pp:swtsr,dls:lecfp} and virial relations~\cite{ca:nmaps,ca:nsbfp}.  The reason for using the FPU model for this
purpose is that it  is one of the simplest and most natural
one-dimensional nonlinear models for
statistical mechanics which can be conceived.  An interesting
alternative of comparable simplicity which
has been the subject of recent research is the truncated Burgers
model~\cite{rva:hssrc,am:smtbh,ajm:rsbtb}.  

Our intent is to use the FPU model to scrutinize a ``weak turbulence''
(WT) theory, a nonequilibrium
statistical mechanical theory which attempts to describe the dynamical
energy transfer among normal modes in a weakly nonlinear, dispersive,
extended system~\cite{djb:nlrwd,jb:rwc,kh:onlet1,bbk:pt,acn:wtaai,acn:wti,vez:kst1}.  The theory has
been 
 developed  over the last four decades to describe the
energy transfer in wave dynamics primarily in fluids and plasmas~\cite{acn:wtaai,acn:wti,vez:kst1}, among other novel applications
such as semiconductors~\cite{yvl:qwtas,yvl:ffsqb,yvl:slks}.

These systems are generally too
complex for effective comparison of the weak turbulence theory with direct
numerical simulations.  Only in recent years has a simple one-dimensional
model with features representative of such fluid systems been
explored by Cai, Majda, McLaughlin, and
Tabak~\cite{dc:sbdwt,dc:dwtod,ajm:odmdw} to examine the
assumptions underlying WT theory.  We propose to apply the FPU model
for a similar purpose, though the issues on which we focus are
distinct.  The implications of our studies for the framework of WT
 theory will be taken up in other works~\cite{prk:awttf}.

Here we will present the content of our findings as they inform the
relaxation process in the FPU model in the  stochastic but
weakly nonlinear regime.  We will restrict attention to the $
\alpha$ version of FPU model, which has purely quadratic nonlinearity
in the equations of motion (Section~\ref{sec:fpu}).  
The $ \beta $ version (with cubic
nonlinearity) seems to be the subject of more work~\cite{ca:nmaps,jdl:ftetl,xl:dame,vvm:cbfce,pp:esfpu,pp:swtsr}, but the $ \alpha $
version has attributes which make it more suitable for a first test
case for WT theory.  Most of the previous statistical mechanical 
work concerning the FPU models of 
which we are aware
focuses primarily on computing particular statistical measures of the
process, such as the time until equipartition is reached
\cite{jdl:ftetl,pp:swtsr}, the
Lyapunov exponents characterizing the degree of
chaos~\cite{lc:fpupr,td:memle,pp:esfpu,dls:lecfp}, or more exotic quantities characterizing the
geometry of the trajectories~\cite{ca:qhpsf,mcs:psgst}.  
Another recent line of research has been tracing the path of energy
transfer starting from a small set of excited
modes~\cite{gc:retip,ky:mioda,ky:preei}. 
Because the WT
theory has the potential to describe the process of
energy transfer in the system  from beginning to end, 
we have instead sought to characterize the
entire evolution of the energy spectrum from large-scale excitation to
eventual equipartition.  We will consider the energy transfer in
spectral terms, in contrast to the physical space viewpoint developed
for the $ \beta$-FPU model by Lichtenberg and coworkers~\cite{vvm:cbfce}.  

The energy spectrum in the $
\alpha$-FPU model approaches equilibrium through a series of
qualitatively distinct phases which we illustrate in Section~\ref{sec:numfpu}.
At the initial time, the energy is concentrated in a small set of
low-wavenumber modes.  This energy then proceeds to higher wavenumbers
first through a standard superharmonic cascade, and then 
shifts to a nonlocal transfer of energy from low wavenumbers
to a band of intermediate wavenumber modes.  The energy in this
intermediate wavenumber band then rolls back through an inverse
cascade to lower wavenumbers again.  This process
then  creates approximate equipartition
only over a set of modes extending up to a cutoff wavenumber, beyond
which the energy content falls off exponentially
rapidly~\cite{ac:tdsrb,lg:peels}.     We refer
to the location of the transition between the flat and rapidly
decaying parts of the energy spectrum as the ``knee.''

After presenting this pictorial ``life history'' of a large-scale
excitation in the $ \alpha$-FPU system, we present in
Section~\ref{sec:scalpred} some specific
quantitative predictions of WT theory and compare them with the
numerical results.  At the coarsest level, WT theory suggests the
presence of two nonlinear time scales.   Over the first nonlinear time
scale, energy is exchanged through triads of modes 
which remain resonant over this time scale.  A
consideration of the resonances in the dispersion relation indicates
that only modes of sufficiently small wavenumber can participate in
nearly resonant triads~\cite{dls:lecfp}.  This in turn suggests that 
this triad interaction phase should correspond to the formation
of the partial equipartition up to the knee.  The subsequent
relaxation of the energy spectrum to global equipartition requires the
slower energy exchange among resonant quartets of normal modes, which are much
more abundant.  Our present
focus is on the triad interaction phase.

An
adaptation of the WT theory allows predictions of both the order of
magnitude of the time scale and the location of the knee which agree
excellently with direct numerical simulation.

\section{Fermi-Pasta-Ulam Model}
\label{sec:fpu}
The Fermi-Pasta-Ulam (FPU) model is a model for a one-dimensional 
collection of particles with
massless, weakly anharmonic (nonlinear) springs connecting them to each other.
Letting $ \{q_j\}_{j=1}^{N} $ and $ \{p_j\}_{j=1}^{N} $ denote the
position and momentum coordinates of an $N$-particle chain, we can
define the FPU model Hamiltonian:
\begin{equation}
H= \sum\limits_{j=1}^N \left(
\frac{p_j^2}{2 m} + \kappa \frac{(q_j-q_{j+1})^2}{2} + \alpha
		\frac{(q_j-q_{j+1})^3}{3}+
                    \beta \frac{(q_j-q_{j+1})^4}{4}
   \right)
\end{equation}
Here we assumed periodic boundary conditions $p_{N+1}=p_1$ and
$q_{N+1}=q_1$.  Equivalently, the beads are connected in a circular
arrangement.  The parameter $ m $ denotes the particle mass, while $
\kappa $, $ \alpha $, and  $\beta $ are coefficients involving the
spring properties.

The equations of motion are the standard Hamilton's equations:
\begin{equation}
\dot{q}_j=\frac{\partial{\cal H}}{\partial p_j}, \ \ 
\dot{p}_j=-\frac{\partial {\cal H}}{\partial q_j}\label{Canonical}
\end{equation}
We will here focus on  the $\alpha$-FPU model  
for which $ \alpha \neq 0 $ and the quartic
term is absent ($ \beta = 0 $).  We nondimensionalize the system
with respect to the spring constant $\kappa $, the mass $ m $, and the
energy density $ H/N $.  Retaining
 the original symbols for the nondimensionalized variables $p_j $ and $ q_j$,
we obtain the nondimensionalized Hamiltonian and equations of motion:
\begin{eqnarray}
\Hamil &=&\frac{1}{2}\sum\limits_{j=1}^N \left( {p_j^2} +
(q_j-q_{j+1})^2\right)+\frac{\epsilon}{3}
\sum\limits_{j=1}^N
(q_j-q_{j+1})^3\label{HamPhys}\\
\dot q_j &=& p_j \nonumber\\
\dot p_j &=& 
\left(q_{j-1}-2 q_j + q_{j+1} \right)\left(1+\epsilon(q_{j-1}-q_{j+1})\right)
\nonumber
\end{eqnarray}
Our choice of nondimensionalization implies that
\begin{equation}
\Hamil = N \label{eq:hamnorm}
\end{equation}
 for all times.
The fundamental nondimensional parameter measuring the strength of the
nonlinearity is
\begin{displaymath}
\epsilon \equiv \alpha \sqrt{\frac{H}{N}}.
\end{displaymath}

In order to study the transfer of energy among different scales, we
represent the system in terms of Fourier modes:
\begin{eqnarray}
\left(\begin{array}{c}q_l \\ p_l \end{array}\right)
=\frac{1}{N}\sum\limits_{k=-\frac{N}{2}+1}^{N/2}
\left(\begin{array}{c}Q_k \\ P_k \end{array}\right)
\exp{\left(\frac{-2\pi i k l }{N}\right)}
\nonumber\\
\left(\begin{array}{c}Q_k \\ P_k \end{array}\right)
=
\sum\limits_{l=1}^{N}
\left(\begin{array}{c}q_l \\ p_l \end{array}\right)
\exp{\left(\frac{2\pi i k l }{N}\right)} 
\end{eqnarray}

The 
Hamiltonian in the new variables reads
\begin{eqnarray}
\Hamil &=& 
\frac{1}{2N}
\sum\limits_{\kdisc=-\frac{N}{2}+1}^{\frac{N}{2}}
\left[ |P_\kdisc|^2+\omega_\kdisc^2 |Q_\kdisc|^2\right] \\
& & \qquad +\frac{\epsilon}{3N^{2}}
\sum\limits_{\kdisca, \kdiscb, \kdiscc= -\frac{N}{2}+1}^{\frac{N}{2}}
\Ifour_{\kdisca, \kdiscb, \kdiscc} 
Q_{\kdisca} Q_{\kdiscb} Q_{\kdiscc}
\delta_{\kdisca+\kdiscb+\kdiscc,0},\nonumber 
\end{eqnarray}
where the dispersion relation is given by
\begin{equation}
\omega_\kdisc=2 \left|\sin{\left(\frac{\pi \kdisc}{N}\right)}\right|,
\end{equation}
the nonlinear coupling coefficients are
\begin{equation}
\Ifour_{\kdisca,\kdiscb,\kdiscc} = -i 
\sgn (\kdisca \kdiscb \kdiscc) \omega_{\kdisca} \omega_{\kdiscb}
\omega_{\kdiscc}, \ \ \ \kdisca+\kdiscb+\kdiscc=0,
\end{equation}
and
\begin{displaymath}
\delta_{i,j} \equiv \begin{cases} 1 & \mbox{if }  i = j \mod N, \\
				  0 & \mbox{else.}
		    \end{cases}
\end{displaymath}
is a periodized version of the Kronecker delta function.

To quantify the amplitude of activity of the FPU chain at different
scales, we define the harmonic energy contribution of each Fourier
mode: 
\begin{displaymath}
\Elinkd (t) \equiv \frac{1}{2N}
\left[ |P_\kdisc|^2+\omega_\kdisc^2 |Q_\kdisc|^2\right].
\end{displaymath}
Energy equipartition implies $\Elinkd$ is independent of $k$ (and
$t$).  The total harmonic contribution to the energy is
\begin{displaymath}
\Elinnd (t) \equiv \sum_{\kdisc = - N/2+1}^{N/2} \Elinkd (t).
\end{displaymath}

\section{Numerical Simulation of Relaxation from
Initial Large-Scale Excitation}
\label{sec:numfpu}
The equations of motion in (\ref{HamPhys}) were integrated using
a fourth order Runge-Kutta time stepping
routine.  Since the equations are local
in physical space, all calculations were performed there.  FFTs were
employed only for data analysis.   All
simulations conserved energy to within $10^{-3}$ after the full
length of the simulation whereas total momentum and particle
position were conserved to machine accuracy.  With a specified number
of excited initial modes, the amplitude and phase of the 
Fourier coefficients are each chosen from a uniform distribution so
that, on average, each mode would be initialized with the same
energy.  The initial data is then normalized according to
Eq.~(\ref{eq:hamnorm}).  
The results of all experiments are
averaged over 10--20 independent realizations of the initial data; 
these relatively small
ensembles proved sufficient to elucidate the results.  

\begin{figure}[tb]
\centerline{
	\includegraphics[width=3in]{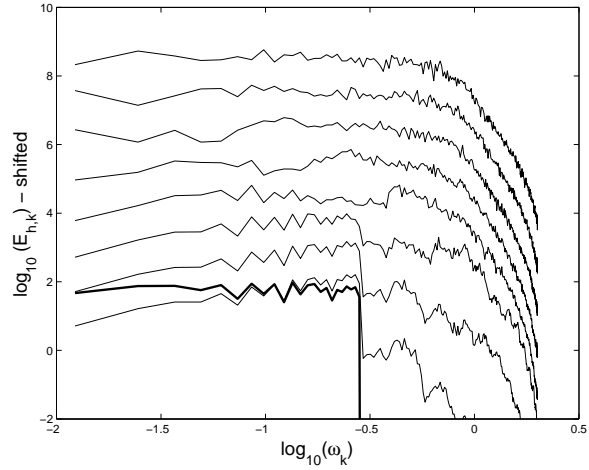}
}
\caption{Temporal evolution of spatial FPU energy spectrum versus
mode frequency for an ensemble of experiments on a lattice of length $
N = 512 $.
Subsequent spectra are shifted upward for ease
of viewing.  Times are, initial (thick line), $t=
50, 100, 200, 400, 1000, 2500, 5000, 10^4$.  Intermediate times
show an inverse cascade whereas late times clearly show a knee,
above which energy decays rapidly. }
\label{fig1}
\end{figure}

Figure \ref{fig1}  shows a sequence of ensemble averaged
spectra for a lattice of length $N=512$ and nonlinearity strength
$\epsilon = 0.03$ at times $t = 0$,
$50$, $100$, $200$, $400$, $1000$, $2500$, $5000$, $10^4$, 
displaced for ease of viewing.
The evolution proceeds from a set of $50$ initially excited modes
to a superharmonic cascade to all wave numbers
with exponentially decreasing energy by $t=50$.  By $t=100$
the initial band has transferred much of its energy to 
intermediate wavenumbers, forming a slight hump. Thereafter, this hump
of energy rolls back via an inverse cascade
 to low wavenumbers.  At $t=10^4$, the energy spectrum exhibits 
a plateau at low wavenumbers and an exponential falloff
at higher wavenumbers.  
This last spectrum is the motivation for
the term ``knee'', below which the waves are in equipartition and
above which they are not substantially
excited~\cite{ac:tdsrb,lg:peels}.  
After $t=10^4$ the
spectrum evolves over much longer time scales, eventually arriving
at equipartition throughout.  

\begin{figure}[tb]
\centerline{
	\includegraphics[width=3in]{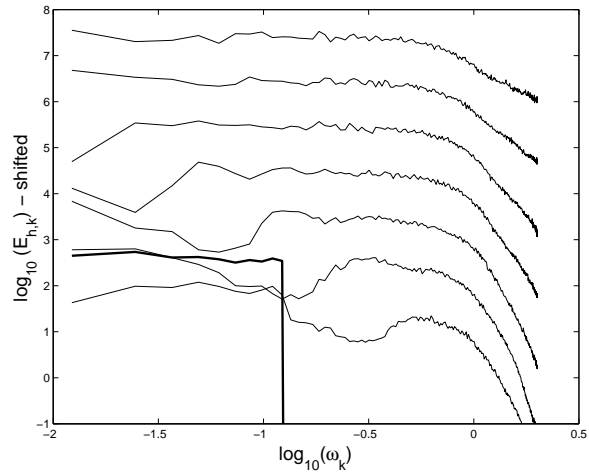}
}
\caption{Temporal evolution of spatial FPU energy spectrum with lattice
length $ N = 512$.  Subsequent spectra are shifted upward for ease
of viewing.  Times are, initial (thick line), 
$t=10^3,2\times 10^3,4\times 10^3,8\times 10^3,
2\times 10^4,5\times 10^4,10^5$
A narrower initial spectrum yields a more pronounced inverse
cascade at intermediate times.  
At late times high wavenumbers begin to acquire more
energy.}
\label{fig2}
\end{figure}

Figure \ref{fig2} shows a similar experiment with a larger ensemble
(20 realizations), $N=512$ and $\epsilon = 0.05$.  
The initial excitation band is very much smaller, including
only 20 waves.  Again spectra are displaced and in this example
are shown at  $t= 0,10^3,2\times 10^3,4\times 10^3,8\times 10^3,
2\times 10^4,5\times 10^4,10^5$.  
Energy is driven first to an intermediate range of wavenumbers which saturate.
Subsequently, an inverse cascade of energy extends the band of
equilibration backward to lower wavenumbers
until  $t \sim 8 \times 10^3$, at which point only the lowest
wavenumber has yet to reach equipartition.  At $t = 5\times 10^4$ the
spectrum is quasi-stationary, equipartition
being achieved among all wavenumbers less than $k_{\mathrm{knee}}$.  The
energy in larger wavenumbers decreases rapidly.  At $t=10^5$
the highest modes begin to acquire energy.  Eventually the
whole spectrum will arrive at equipartition;  this process
is outside of the scope of the current work.  

\section{Scaling Predictions}
\label{sec:scalpred}
A renormalized WT theory derived in~\cite{prk:awttf}
predicts that significant three wave interaction should occur
in a  band 
\begin{equation}
|k| \le k_{\mathrm{knee}} \sim N \sqrt{\epsilon} \label{eq:kscale}
\end{equation}
on a time
scale
\begin{equation}
 T_{3} \sim \epsilon^{-3/2}. \label{eq:Tscale}
\end{equation}
This theory only applies when the lattice size is large enough ($ N
\gg \epsilon^{-1/2} $) and the number of initially excited modes is
an order unity fraction of the knee width $ \kknee $, so that the
renormalized energy spectrum remains self-consistently of order unity
during this phase of evolution.

\begin{figure}[tb]
\centerline{
	\includegraphics[width=3in]{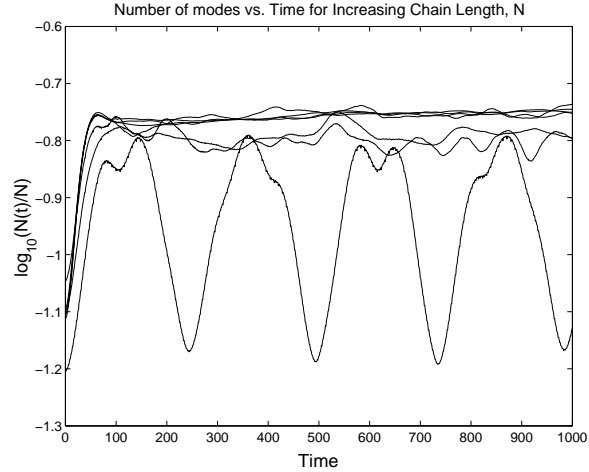}
}
\caption{$\log_{10}(\neff(t)/N)$ versus time
for lattice lengths $N=32, 64, 128, 256, 1024, 2048, 4096$, $\epsilon = 0.1$.
The simulations with large lattice length lie along one another
in confirmation of the universal scaling predictions (\ref{eq:kscale})
and (\ref{eq:Tscale}).  Lattices with $N\le 128$
 show some mild quasi-periodicity whereas the 
simulation $N=32$ clearly shows the quasi-periodic
behavior of the original FPU simulations~\cite{ef:snp}.}
\label{fig5}
\end{figure}

A useful statistical measure for our purposes is 
the spectral entropy, defined as
\begin{displaymath}
S (t)  = -\sum_k \frac{E_{h,k}(t)}{E_h(t)} 
\log\left(\frac{E_{h,k}(t)}{E_h(t)}\right).
\end{displaymath}
This provides a measure of the effective 
number of excited normal modes at any given
time, $\neff (t) \equiv
e^{S(t)}$~\cite{lc:fpupr,jdl:ueeoc,rl:etnlh,vvm:cbfce,pp:swtsr}. 
Figure \ref{fig5} shows rescaled plots of this spectral 
entropy as a function of time.  
The onset of the quasi-stationary phase, 
after the end of the three-wave evolution, is clearly
evident.  The knee width $k_{\mathrm{knee}} \approx
 1.5 N \sqrt{\epsilon}$ is determined as an average of $ \neff $ 
over a time window shortly after the entropy ends its rapid rise.
This scaling relationship is robust against various choices of initial
bandwidth excitations.  The time to reach partial equipartition $ T_3 $,
however, does depend sensitively on the choice of initial data.  As
discussed above, the
WT theory producing the scaling prediction
(\ref{eq:Tscale}) assumes the initial data is excited over a band of
wavenumbers which is an order unity fraction of the knee width.  To
test the prediction (\ref{eq:Tscale}), then, we choose the system to
have initially $ \frac{1}{2} \kknee \approx 0.75 N \sqrt{\epsilon} $
excited modes.  (The evolution depicted in Figure~\ref{fig1} comes
from initial data of this form, whereas the example in
Figure~\ref{fig2} was initialized with a considerably smaller band of
excited modes). 
The time scale $ T_3 $ for the system to reach partial
equipartition is determined automatically as the first time at which
$ \neff (t) $  achieves the value $ \kknee = 1.5 N \sqrt{\epsilon}$.

\subsection{Effect of strength of nonlinearity}

\begin{figure}[tb]
\centerline{
	\includegraphics[width=3in]{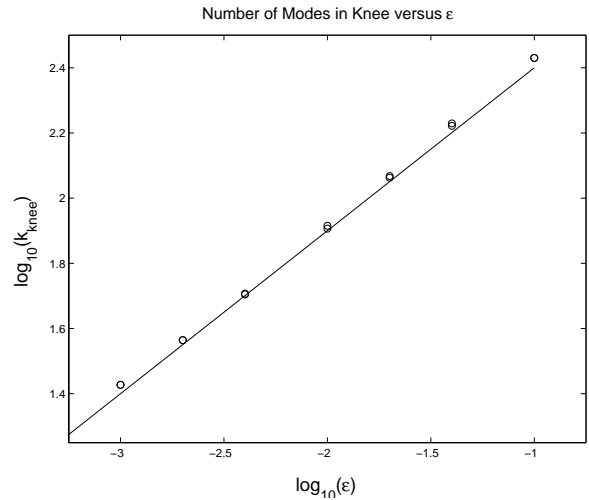}
}
\caption{Knee width, 
$\log_{10}(k_{\mathrm{knee}})$ versus $\log_{10}(\epsilon)$ for $N=512$
and initial data chosen at half the predicted knee width. The line
represents the scaling law $ \kknee = 1.5 \epsilon^{1/2} N$.}
\label{fig3}
\end{figure}

\begin{figure}[tb]
\centerline{
	\includegraphics[width=3in]{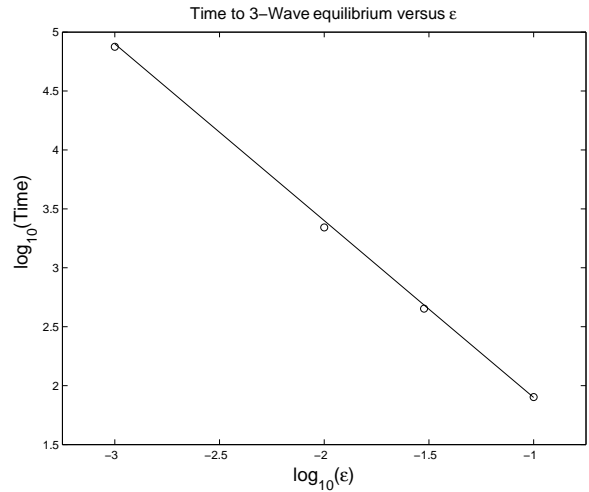}
}
\caption{Time to three-wave equilibrium,
$\log_{10}(T_{3})$ versus $\log_{10}(\epsilon)$ for $N=512$
and initial data chosen at half the predicted knee width. The line
represents the scaling law $ T_3 = 2.5 \varepsilon^{-3/2} $.}
\label{fig4}
\end{figure}

Ensembles of FPU lattices with lattice length $ N = 512 $ were
 integrated for values of $\epsilon$ ranging from $10^{-3}$ to $10^{-1}$.
Figures \ref{fig3} and \ref{fig4}
show excellent agreement with the $ \epsilon $ scaling dependences
(\ref{eq:kscale}) and (\ref{eq:Tscale}) 
predicted by the renormalized WT theory.  The scaling $
\kknee \sim \epsilon^{1/2} $ had also been previously observed
in~\cite{dls:lecfp}.  

\subsection{Effect of lattice size}
The renormalized WT theory also predicts that the
knee width should scale with lattice size (Eq.~(\ref{eq:kscale})) whereas the
three wave time scale should not (Eq.~(\ref{eq:Tscale})).  These
properties are
compatible with a thermodynamic limit.  
In figure \ref{fig5} the logarithm of the
fraction of modes to the total lattice size 
is plotted against time for seven ensembles of experiments with
increasing lattice size $N=32, 64, 128, 256, 1024, 2048, 4096$,
and fixed $\epsilon = 0.1$.  
These experiments were again initialized with half
the number of modes of the predicted knee.
Three features of this plot
stand out most clearly.  First, the spectral entropy follows a universal
evolution~\cite{jdl:ueeoc,pp:swtsr} for 
lattice sizes larger than
$N=128$. Secondly, the 
number of excited modes  exponentially increases with time
prior to three wave equilibrium.  Finally, for $N=32$ (where the
initial number of excited modes is approximately $7$) there
is no equilibrium, but rather quasi-periodic behavior.  In fact,
a shadow of this behavior is present for $N=64$ and $128$ also.  
These are reminiscent of the integrability discovered in 
Fermi, Pasta and Ulam's original work~\cite{ef:snp}, and indicate the
breakdown of the WT theory scaling predictions for small
lattice sizes.

\section{Conclusion}
We have emphasized some of the long-lived transient features of the
evolution of the energy spectrum in the $ \alpha$-FPU model.  Weak
turbulence theory has been successful in predicting scaling
exponents concerning the achievement of partial equipartition.
In the future, we will endeavor to explain other dynamical aspects,
such as the formation of the energy hump at intermediate wavenumbers and
subsequent inverse cascade, in similar quantitative terms.  

\section{Acknowledgements}
The authors would like to thank David Cai and Gregor Kova\v ci\v c for
helpful discussions.  JAB
is supported by an NSF VIGRE postdoctoral research fellowship DMS
9983646, 
PRK is supported by an NSF grant DMS-A11271, 
and YVL
is partially supported by NSF Career grant DMS 0134955 and
ONR YIP grant N000140210528.  

\bibliographystyle{plain}

\bibliography{journalorig,journalmy,publishorig,publishmy,weakturb,statphys,waves,own}

\end{document}